\documentclass[traditabstract]{aa} 
\usepackage{graphicx}
\usepackage{subfigure}
\usepackage{amssymb, amsmath}
\usepackage{subfigmat}                     
\usepackage{amsmath}
\usepackage{graphicx}
\usepackage{ulem}
\usepackage{amsmath}
\usepackage{graphicx}
\usepackage{url}
\usepackage[utf8]{inputenc}

\newcommand{\beq}{\begin{equation}}
\newcommand{\eeq}{\end{equation}}
\newcommand{\bea}{\begin{eqnarray}}
\newcommand{\eea}{\end{eqnarray}}
 \makeatletter

\def\@biblabel#1{\hspace*{-\labelsep}}

\makeatother 

\begin{document}

\title{Small-scale dynamos on the solar surface: dependence on magnetic Prandtl number}
\titlerunning{Small scale dynamo action at low magnetic Prandtl numbers?}

\author{Irina Thaler \inst{1}  \inst{2}  H.C.\ Spruit \inst{1}}
\authorrunning{I.\ Thaler \& H.C.\ Spruit}
\institute{Max-Planck-Institut f\"{u}r Astrophysik,
  Karl-Schwarzschild-Str.\ 1,
  D-85748 Garching, Germany \and
Kiepenheuer Institute for Solar Physics,
Sch\"{o}neckstra{\ss}e \ 6, D-79104 Freiburg im Breisgau, Germany  }

 \date{\today}
 \abstract{The question of possible small-scale dynamo action in the surface layers 
 of the Sun is revisited with realistic 3-D MHD simulations. As in other 
 MHD problems, dynamo action is found to be a sensitive function of the 
 magnetic Prandtl number ${\rm P_{\rm m} }=\nu/\eta$; it disappears below a critical 
value ${\rm P_{\rm c}}$ which is a function of the numerical resolution. At a 
grid spacing of 3.5 km, ${\rm P_{\rm c}}$ based on the hyperdiffusivities 
implemented in the code (STAGGER) is $\approx 1$, increasing with 
increasing grid spacing. As in other settings, it remains uncertain 
whether small scale dynamo action is present in the astrophysical limit 
where ${\rm P_{\rm m} }<<1$ and magnetic Reynolds number ${\rm R_m}\gg 1$. The question is 
discussed in the context of the strong effect that external stray fields 
are observed to have in generating and maintaining dynamo action in 
other numerical and laboratory systems, and in connection with the 
type-II hypertransient behavior of dynamo action observed in the absence 
of such external fields.}
\maketitle

\section{Introduction}

In laboratory experiments and numerical simulations of hydrodynamic turbulence, it is found that at sufficiently high Reynolds numbers the statistical properties of the flow become independent of the viscosity value ($\nu$). This observation has been enshrined  in the standard  cascade picture for three-dimensional (but not two-dimensional) turbulence (Kolmogorov 1941).  The addition of a magnetic field introduces a second dissipative mechanism, the magnetic diffusivity $\eta$. By analogy with the hydrodynamics case, it may be assumed that a turbulent magnetic flow would also be insensitive to the value of $\eta$, such that it would only affect the small scales in the magnetic field. Experience with numerical magnetohydrodynamical (MHD) simulations has shown this assumption to be unexpectedly problematic, however. The behavior of turbulent MHD flows, including their bulk transport efficiency and the presence or absence of small-scale dynamo action, appears to vary with details of the problem studied, with the numerical methods used, and in particular appears to depend on the magnetic Prandtl number of the fluid, ${\rm P_{\rm m}} =\nu/\eta={\rm R_{m}/R_{e}}$. 

The magnetic field at the surface of the Sun is dominated by the sunspot cycle, which is believed to have its source near the base of the convective envelope. At sufficient spatial resolution, a weak field of mixed polarity is also observed. Although it is not clear from the observations whether this is a truly separate component and not some sort of waste product of the spot cycle, it  suggests that a local small-scale turbulent dynamo process might be operating
 in the surface layers (Durney et al.\ 1993; Petrovay \& Szakaly 1993). This suggestion has been addressed with a number of more idealized turbulent dynamo models and simulations (Cattaneo 1999; Cattaneo et al.\ 2003). In view of the poorly understood dependence of MHD turbulence on details of the problem studied, these simulations do not yield an unambiguous interpretation of the weak field observed on the Sun. With the advent and spectacular successes of realistic 3D radiative MHD simulations developed for the solar surface layers (Galsgaard \& Nordlund 1996), it has become possible to study the problem numerically for conditions much closer to the solar case, where V{\"o}gler \& Sch{\"u}ssler (2007), Pietarila Graham et al.\ (2010)  and Rempel (2014)  reported successful small-scale dynamo generation with such simulations. V{\"o}gler \& Sch{\"u}ssler (2007) and Pietarila Graham et al.\ (2010) investigated a range in magnetic  Prandtl numbers and Reynolds ${\rm R_m}$, but  results were still somewhat inconclusive for the combination of large ${\rm R_m}$ and small ${\rm P_{\rm m}}$. We here follow up on these results, with emphasis on the dependence on ${\rm P_{\rm m}}$ at high resolution.
\section{Small-scale dynamos  at low $\mathbf {P_{\rm m}}$ }
A very similar dependence on the magnetic Prandtl number has been observed in a variety of physical conditions where small-scale self-sustained magnetic  field generation (\`dynamo action') is expected. These conditions range from idealized models to  astrophysical accretion disks and solar surface convection. In all these cases the dependence has been discovered through numerical simulations, the results of which appeared to contradict basic notions of self-sustained field generation. \\
That macroscopic behavior of MHD might depend critically on the magnetic Prandtl number was previously noted by Balbus and Hawley (1998) in the context of small-scale dynamo action in accretion disks. An often-used model for small-scale dynamo action is the so-called fluctuating dynamo:  a flow driven by an assumed  external force acting on a large scale and with a random time dependence imposed on it. It is intended to be generically applicable to MHD turbulence, and believed to be sufficient to prove dynamo action (but not without controversy, see references in Iskakov et al.\ (2007)) \footnote{For large-scale (global) dynamo action, the differences between turbulence models and numerical results appear to be even more contentious. See, for example, Hughes et al.\  (2010) and references therein.}. Numerical simulations (Schekochihin et al.\ 2002) showed, however, that the presence of self-sustained magnetic field generation in this model depends critically on the magnetic  Prandtl number, with dynamo action absent when ${\rm P_{\rm m}}  \la 1$ (viscosity lower than magnetic diffusivity). The magnetic Prandtl number divides astrophysical systems into two very different regimes. It is usually either very small, as in the case of stellar convective  cores,  or very large, as in the interstellar or intracluster medium (Schekochihin et al.\ 2007).  

At higher Reynolds numbers or numerical resolution, Iskakov et al.\ (2007) found dynamo action below the explicit numerical magnetic Prandtl number $ \rm{ P_{\rm m}=1}$  
in the fluctuating model, at low growth rates. The growth rate increases with magnetic Reynolds number, but at the lowest ${\rm P_{\rm m}}$ 
 achieved, $\approx 0.1$, it was still declining with decreasing
${\rm P_{\rm m}}$. 
 The authors proposed that the results indicate a  flattening of the growth rate to a finite positive value for ${\rm P_{\rm m}}  \downarrow 0$.   Functions that decline continuously to zero, would equally fit the data shown in their Fig.\ 1, however.  
 Using a different assumption about the driving velocity field, Boldyrev \& Cattaneo (2004) proposed that dynamo action remains possible at any low value of   ${\rm P_{\rm m}}$, at sufficiently high ${\rm R_{\rm m}}$.  Each of these models incorporates assumptions  of uncertain physical realism about the small-scale velocity field, however, and does not yield unambiguous results for  the scaling with  ${\rm P_{\rm m}}$  of the minimum ${\rm R_{\rm m}}$ required for dynamo action.
The existence of small-scale dynamo action therefore does not appear to be settled for astrophysically relevant magnetic Prandtl and Reynolds numbers. 

A physically realistic yet simple model that does not need an assumed small-scale velocity field or  forcing  is the shearing box model for the flow in an accretion disk. Strong magnetorotational turbulence develops in rotating shear flows that are stable in the absence of a magnetic field  (such as Keplerian shear). Lesur $\&$ Longaretti (2007) and Fromang et al.\ (2007), see also Fromang (2010, 2010b),  and  Simon (2011) studied the dependence of magnetorotational turbulence  on  ${\rm P_{\rm m}}$  through simulations in which they explicitly included viscosity and magnetic diffusivity.   The results show that the amount of transported angular momentum increases with  ${\rm P_{\rm m}}$.  The dependence on  ${\rm P_{\rm m}}$  is  similar to that found in the fluctuating dynamo simulations (Riols et al.\ 2014).  The auhors found that the presence or absence of a (weak) mean vertical field (a net flux crossing the disk,  see also Sect.\ \ref{weakf}) is a crucial factor.

Since most codes used in astrophysics do not include explicit viscosity or magnetic diffusivity, but leave this to the discretization errors   and  the stabilization algorithms of the code, the value of  ${\rm P_{\rm m}}$ that is to be associated with a  particular  simulation is not obvious  a priori.  The action of these algorithms is not equivalent to physically realistic diffusion coefficients. If the algorithm used for the stabilization term (such as a hyperdiffusivity) is of the same form for the momentum and magnetic field vectors, however, one would expect its effect to correspond to a Prandtl number around unity. By varying the relative amplitude of the stabilization terms in the momentum and induction equations, the effect of varying the magnetic Prandtl number can be mimicked.  \\

 In most simulations such adjustments are not made. This raises the question which value of the magnetic Prandtl number best characterizes the numerics. This can be studied by measuring effective transport coefficients  from the simulations, for example, from the evolution of additional 
vectors for momentum and magnetic field that are advected in the same way, but do not contribute to the equations of motion and induction (so-called ghost vectors).  The value derived by Fromang et al.\  (2009)  is  ${\rm P_{\rm mn}}\approx 2;$ 
similar numbers have been reported for other codes.  Here ${\rm P_{\rm mn}}$ indicates the Prandtl-like ratio of numerical diffusivities to keep a clear distinction from the physical magnetic Prandtl number ${\rm P_{\rm m}}$.  
This value, in the range where dynamo action is also observed in simulations with explicit diffusivities, explains  why  magnetorotational turbulence has been obtained in most astrophysical codes. For the results reported below, the code includes an explicit process for modifying its effective value of the numerical magnetic Prandtl number  ${\rm P_{\rm mn}}$. 

The reason for the strong dependence on  ${\rm P_{\rm m}}$  has been discussed in terms of  the ordering of the viscous and  resistive length scales (cf.\ Moffatt (1961)). For  ${\rm P_{\rm m}}\gg 1$ , the viscous length scale, where the field stretching takes place, is much larger than the resistive one, which  then plays  a negligible role (Batchelor 1950). The situation is quite different for  ${\rm P_{\rm m}} \ll 1$, when the resistive scale is much larger than the viscous scale.  As argued in  Schekochihin et al.\ (2004b), in this case, small scales in the field dissipate faster than they can be amplified by field line stretching in the viscous flow.  The authors also noted that in this interpretation, self-sustained dynamo action is not achievable in liquid metal laboratory experiments, because the required magnetic Reynolds numbers would be far beyond experimental feasibility.

\subsection{Transient behavior}
\label{transi}

A remarkable observation that might also be relevant for the solar case has been made by E. Rempel et al. (2010), who studied the classical shearing box model of magnetorational dynamo action at  ${\rm P_{\rm m}}>1$   
 for the case where the mean field through the disk vanishes.  Dynamo action saturates rapidly (a few orbital timescales) to a statistically steady state, but after a finite time switches off again on an equally fast timescale. Since onset of dynamo action requires a finite seed field to overcome magnetic diffusion, this inactive state is final.
 This type of behavior has been observed in other chaotic systems and is called \textbf{{\it type-II supertransient}} behavior.  The duration of the active phase increases approximately   \textbf{{\it exponentially}}  with increasing ${\rm R_m}$.  The switch-off is explained as occurring when decay-facilitating fluctuations are  by chance simultaneously present  in all statistically independent subvolumes of the simulation. This occurs more readily at low spatial resolution or Reynolds number and explains the exponential dependence of the transient duration on ${\rm R_m}$.  \\
 Whether the supertransient phenomenon is relevant for stars is not clear, however. On the one hand, the steep increase of the transient duration with Reynolds number argues against  a possible relevance in environments with a high Reynolds number; on the other hand, some relevant timescales are also very long in stars. 

In the presence of a finite net flux that threads the disk,  the supertransient  effect cannot take place, since this flux cannot change (cf.\ Spruit \& Uzdenksy (2005) for net flux in accretion disks).  Provided the strength of this net flux   exceeds the minimum required for the magnetorotational instability to grow, it  can  restart  the dynamo process when it is on its way to  decay in a supertransient accident. The result would be a large fluctuation instead of a switch-off.  The key in this case is the conservation property of a net flux crossing a disk. The flux of an initial field consisting of loops that are closed within the box, for example, or a disk-crossing flux consisting of canceling polarities, is not conserved. It does not have the same effect as a net flux; it just acts like an arbitrary initial condition whose imprint vanishes with time. 

A generalization of the net flux crossing an accretion disk is possible. A mean field that can be considered as being imposed externally, for some reason, would have the same effect, for example, a large-scale field produced by an independent dynamo process (cf. Sect.\ \ref{weakf}). To conclude these considerations,  the presence of some low level of magnetic flux from an external source may be essential for the long-term stability  of small-scale dynamo action at any value of the magnetic Prandtl number  (see also the speculations in Sect.\ \ref{weakf}).

\subsection{Experimental evidence}
Liquid metals have low magnetic Prandtl numbers (on the order of $10^{-5}$) that are conveniently in the astrophysically relevant range. Reaching the high Reynolds numbers expected to be needed for dynamo action has been more challenging. The most successful experiment so far has been reported by Monchaux et al.\ (2007). In a turbulent shearing flow between counter-rotating plates in liquid sodium, these authors obtained dynamo action at ${\rm R_m}\approx 50$ in the form of a steady field with superposed strong fluctuations. This success appeared to be related to a peculiarity of the experimental device. Dynamo action was absent until one of the rotating parts, made of stainless steel (low magnetic permeability), was replaced by an iron part (high permeability and magnetic remanence). The remanent magnetization of the iron part may have played a role. Once magnetized by the steady component of the magnetic field, the iron part would have maintained a minimum field strength in its neighborhood. This is significant in view of the experience (for example in the shearing box simulations mentioned above) that even a weak externally imposed field component strongly facilitates dynamo action. Whatever the precise interpretation, however,  the relevance of this experimental result for astrophysics is questionable since astrophysical fluids with the magnetic properties of solid iron are unknown.

\subsection{Situation for the Sun}
 How much the dynamo mechanism depends on the details of the physics included is under debate ( e.g., Moll et al.\ (2011) and references therein). A range of models has been used to study local solar dynamo action, reaching from Bossinesq models (Cattaneo 1999, 2003) and compressible convective models in polytropic layers (Brummell et al.\ 2010; Bushby et al.\ 2010, 2012; Bushby \& Favier 2014) to realistic solar surface models (Stein 2003; V{\"o}gler \& Sch{\"u}ssler 2007; Pietarila Graham 2010; Rempel 2014).
Bushby \& Favier (2014) found that the critical magnetic Reynolds number increases with decreasing magnetic Prandtl numbers; they achieved a minimum value of  ${\rm P_{\rm mn} \approx 0.1}$.
 In realistic 3D MHD solar surface simulations, magnetic Prandtl numbers  ${\rm P_{\rm mn}} \approx 1$  are accessible, and for these values small-scale dynamo action has been found
(V\"ogler \& Sch{\"u}ssler 2007; Pietarila Graham et al.\ 2010;  Rempel 2014). 
In terms of dynamo behavior, 
  ${\rm P_{\rm m}}= 1$  still belongs to the large magnetic Prandtl number limit, however. In view of the inconclusive results discussed above,  the question is still open whether low- ${\rm P_{\rm m}}$  small-scale dynamo action  is to be expected on the solar surface. 

\subsubsection{Indications of solar small-scale dynamo action}
Livingston \& Harvey (1971) discovered an intrinsically weak small-scale internetwork field on the Sun.  It is spread approximately uniformly across the solar disk and seems to be independent of the solar cycle. Its properties have been studied in detail by Martin (1988, 1990) and Zirin (1985). Durney et al.\ (1993) and Petrovay \& Szakaly (1993) suggested that this component might be due to  
small-scale dynamo action located near the solar surface. Alternatively, the weak-field component might represent fragments of active regions rising through the convection zone, or be a by-product of the decay of active regions (e.g.\ Spruit et al.\ (1987)). In this decay hypothesis a correlation between the quiet-Sun magnetic field and the solar cycle would be expected. This is not evident in the observations, therefore it would require that the decay from the 
large-scale magnetic field to the smallest scales exceeds the solar cycle time, in this interpretation (Lites 2011).  On the other hand, Parnell et al.\ (2009) found that the magnetic flux distribution between $10^{17}-10^{23}$ Mx can be described by a single power-law function. This would indicate that the whole field is produced by the same process.
Since weak fields tend to be compressed into strong fields by the granulation flow, (some fraction of) the intrinsically strong small-scale magnetic field might be unrelated to the sunspot cycle, but instead 
 result from a small-scale dynamo mechanism.
The origin and possible variation of the strong-field component is of special interest because of its brightness contribution to the total solar irradiance (Schnerr $\&$ Spruit 2011;
Foukal et al.\ 2006; Afram et al.\ 2011; Thaler $\&$ Spruit 2014).

\section{Calculations}
\subsection{Numerical methods}
 The numerical simulations were realized with the 3D 
 magnetohydrodynamics code STAGGER  (Nordlund \& Galsgaard 1995;
 Galsgaard \& Nordlund 1996).  The  code solves the time-depended
 magnetohydrodynamics equations by a sixth-order finite-difference scheme using fifth-order interpolations for the spatial
 derivatives, while the  time evolution is calculated using a third order
 Runge-Kutta scheme. For every time step the radiative transfer
 equation is solved at every grid point assuming local thermal
 equilibrium. This is done by using a Feautrier-like scheme along the
 rays with two $\mu$ angles plus the vertical and four $\phi$ angles
 horizontally, which adds up to nine angles in total.
To incorporate the wavelength dependence of the absorption
coefficient, the Planck function is sorted into four opacity bins.
The equation-of-state table is calculated using a standard program for ionization equilibria and absorption coefficients 
(Gustafsson et al.\ 1973)
 and using opacity distribution functions identical to those used
 by Gustafsson et al.\ (1975), which are described in more detail in Stein \& Nordlund
 (1998), Nordlund (1982) and Nordlund \& Stein (1990).  For a more detailed 
 description of the STAGGER code see Nordlund \& Galsgaard (1995),  and 
 Beeck et al.\ (2012).
 The horizontal boundaries are periodic, while at the top and bottom are open transmitting boundaries.  The effective surface temperature resulting from the model is controlled through the entropy value  of gas flowing in at the 
lower boundary, which is kept fixed during the simulation.  The magnetic field is kept
vertical at the lower boundary, thus allowing horizontal displacements of the field lines. At the top boundary the horizontal field components are determined from the vertical component by a potential field extrapolation.

\subsection{Setup}\label{sec:set-up}
The setup we used consists of a box with  horizontal dimensions of 3 Mm $\times$ 3 Mm and 1.3 Mm vertical, extending to 475 km above the photosphere and 836 km below. The initial magnetic field configuration consists of a checkerboard pattern with a vertical magnetic field of alternating polarity  and strength of 10 mG or
1 mG. This is similar to the setup used in V{\"o}gler \& Sch{\"u}ssler (2007) and Pietarila Graham et al.\ (2010), who used a 4 $\times$ 4 checkerboard in a  box  with horizontal dimensions of  6 Mm  $\times$ 6 Mm
 with a vertical depth of 1.4 Mm in parts of their experiments.

\subsection{Implementation of the  numeric  magnetic Prandtl parameter  $ {\rm P_{\rm mn}}$  in the STAGGER code}
 In the STAGGER code the stabilization  at small length scales is implemented through hyperdiffusivities.  As already mentioned,  the numerical magnetic Prandtl number  $ {\rm P_{\rm mn}}$  is  defined as the ratio between the magnetic hyperdiffusivity and the numerical hyperviscosity. 
Both of them are implemented following the same scheme, but by using a different prefactor, a variation in  ${\rm P_{\rm mn}}$  can be produced. A detailed description of the diffusion scheme can be found in Nordlund \& Galsgaard (1995) (\url{http://www.astro.ku.dk/~kg/Papers/MHD_code.ps.gz})  and Stein and Nordlund (1998). A similar procedure is used in other solar surface convection codes to implement hyperviscosity, which can be compared qualitatively but not quantitively with the procedure from the STAGGER code (V{\"o}gler et al.\ 2005; Pietarila Graham et al.\ 2010; Rempel et al.\ 2009; Rempel 2014), and therefore the derived magnetic Prandtl numbers cannot be directly compared either.

\section{Results}
\begin{figure}                                          
\vspace{-3cm}
\includegraphics[width=0.5\textwidth]{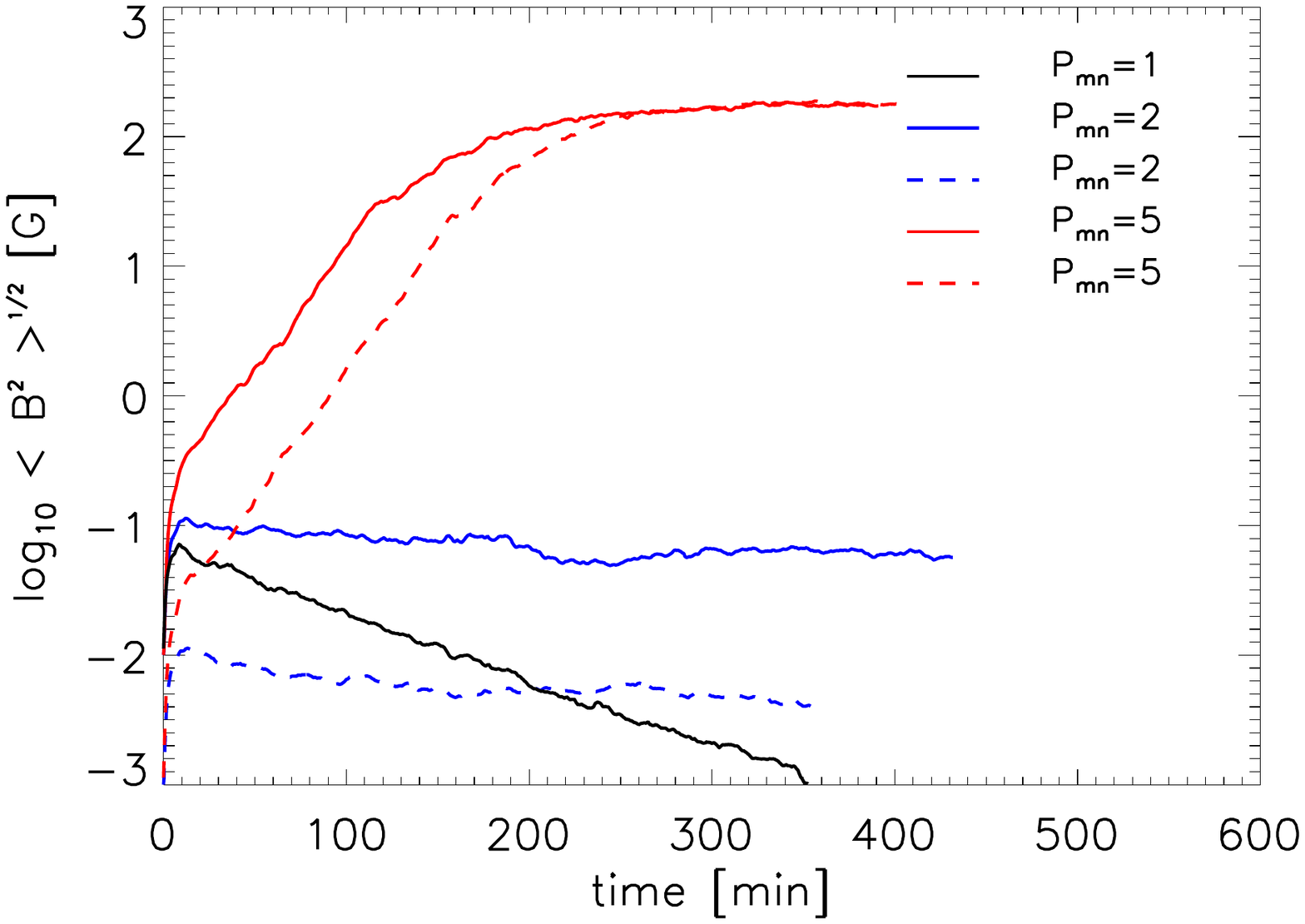}
\vspace{-3cm}
\caption{ Evolution of $\langle B^2\rangle ^{1/2}$ as a function of the numerical magnetic Prandtl parameter  ${\rm P_{\rm mn}}$  for a numerical resolution of $\Delta x= \rm 7$ km and $\Delta z_{0}=\rm 6$ km. The solid line denotes the initial field strength of 10 mG, the dashed line the strength of 1 mG.}
\label{fig:P_mn_dependency}

\end{figure}

A set of simulations  was made to investigate the dependence  of magnetic field amplification on the numerical magnetic Prandtl parameter  $ {\rm P_{\rm mn}}$. The dependence on numerical resolution, as a proxy for dependence on Reynolds number was investigated with an additional set of three runs. Figure \ref{fig:P_mn_dependency} shows the dependence on  ${\rm P_{\rm mn}}$. The boundary between presence and absence of dynamo action appears to be close to  ${\rm P_{\rm mn}=2}$ . 
For ${\rm P_{\rm mn}=5}$  the  amplitude of the field increases exponentially with a growth time of $\sim 200$ min, saturating at $\langle B^2\rangle^{1/2}\approx 115$ G. 

\begin{figure}                                        
\includegraphics[width=0.5\textwidth]{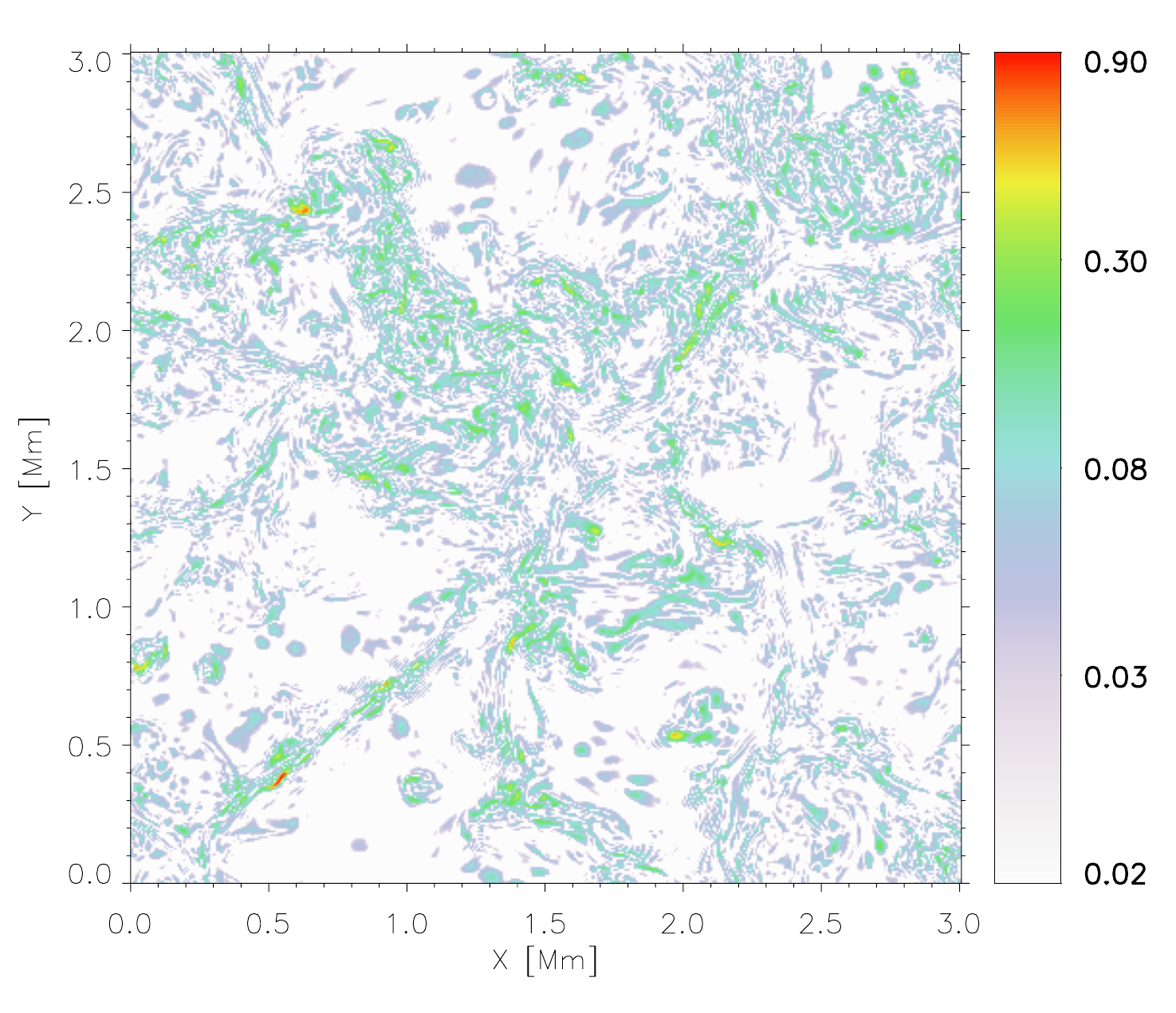}
\caption{ $|B_{z}|$ in kG at the photosphere for  ${\rm P_{\rm mn}=5}$  after t=231 min for a numerical resolution of $\Delta x= \rm 7$ km and $\Delta z_{0}=\rm 6$ km. The image scale is exponential ($val^{0.25}$) %
 to amplify small structures and still be able to clearly identify high values}
\label{fig:Bz_phot}
\end{figure}
\begin{figure}[h!]                                          
\includegraphics[width=0.5\textwidth]{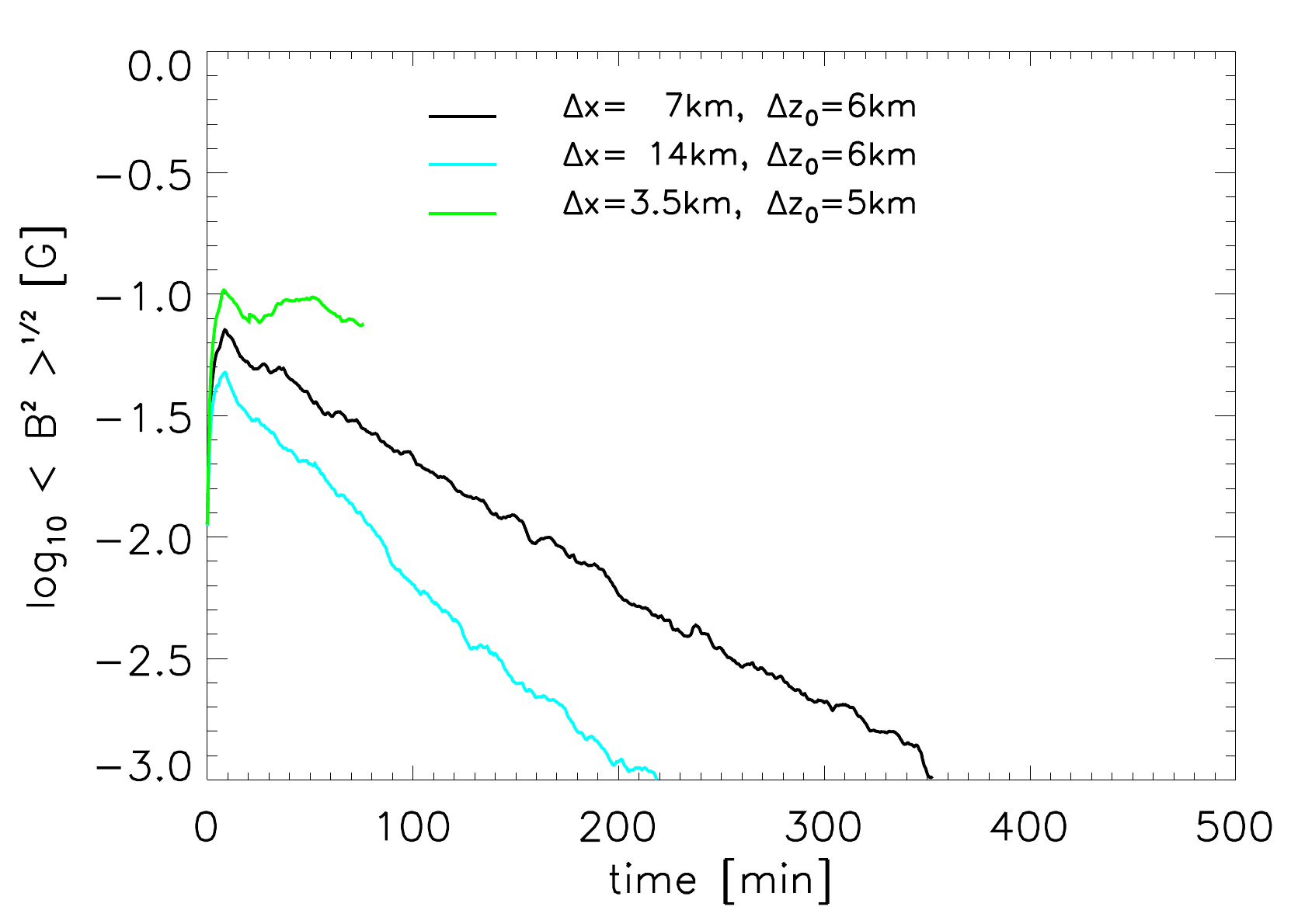}
\caption{ Evolution of $\langle B^2\rangle^{1/2}$ as a function of numerical resolution, for  $P_{\rm mn}=1$  and an initial field strength of 10 mG.}
\label{fig:P_mn_1_res_dependency}
\end{figure}

 
Figure \ref{fig:P_mn_1_res_dependency} shows simulation runs for an initial state
of 10 mG and a numerical Prandtl parameter  ${\rm P_{\rm mn}=1}$   for different numerical
resolutions.  After a short amplification phase of about 10 min, a decline of the magnetic energy sets in.
The code uses a uniform  resolution in the two horizontal coordinates  ($\Delta x$ in Fig. \ref{fig:P_mn_1_res_dependency}); the vertical resolution is non-uniform. It is chosen highest near the photosphere where cooling by radiation drives the flows (grid spacing at the photosphere is denoted by $\Delta z_0$ in the figure).  The figure shows that the decay time increases with increasing resolution, from about 50 minutes at $\Delta x=14$ km to 70 min at 7 km.
At the (computationally expensive) resolution of 3.5 km the limited time coverage of the run indicates decay at an even slower rate. Extrapolation to even higher resolution is uncertain from these data, however.

Since the dissipation built into the code decreases with decreasing grid spacing, the inverse of resolution is a numerical analog of a Reynolds number. Since the nature of the numerical dissipation is rather different from physical dissipation and strongly depends on the order of the spatial discretization used, however, it is not possible to translate the grid spacing in our simulations meaningfully into effective magnetic Reynolds numbers. The uncertainty is smaller for the magnetic Prandtl number, since it measures the ratio of two quantities which, although both artificial, are based on the same algorithm.

\section{Discussion and conclusions} 
\label{discu}

The results show the same trends as were found in the shearing box simulations discussed in the introduction: the growth rate of dynamo action increases with magnetic Prandtl number, and  no dynamo action is detected below a critical value  ${\rm P_{\rm mn}}\approx 1$. Figure \ref{fig:P_mn_1_res_dependency} showed that the growth rate appears to increase with numerical resolution, however, and the corresponding critical  $ {\rm P_{\rm mn}}$  may increase as well. The combination of high resolution and low ${\rm P_{\rm mn}}$ required to study this is numerically challenging.

The growth rate we derived for ${\rm P_{\rm mn}=5}$ is comparable with  the rate reported by Pietarila Graham et al.\ (2010) for their ${\rm P_{\rm mn_{ eff}}}\approx 2$. 
Since the numerical Prandtl parameter ${\rm P_{\rm mn}}$ is not identical
with the physical Prandtl number  ${\rm P_{\rm m}}$, but depends on the
implementation of energy dissipation at small scales in the code, a
dependence of the effective Prandtl number scale on the code used was to be expected.

Assuming a difference of a factor of  a few  between the scales, the lowest  ${\rm P_{\rm mn_{ eff}}}$  $\approx 1$ reported in Pietarila Graham et al.\ (2010) would correspond to  a few  ${\rm P_{\rm mn}}$  in our simulations.
The fact that all results reported by Pietarila Graham et al.\ (2010) showed dynamo action is thus consistent with our results. 

Taken together, these results clearly show that the question of small-scale dynamo action in the surface layers of the Sun is still unresolved. As for the fluctuating dynamo model discussed in the introduction, the limit of a high magnetic Reynolds number (or numerical resolution) combined with a low Prandtl number cannot be reliably extrapolated from currently available results.
\subsection{Relation between intrinsically weak and strong surface fields}
Given that the small-scale dynamo acts and produces the weak-field
component, might (some fraction of) the intrinsically strong component originate from
it as well? 
This seems unlikely from the results obtained with our simulations. 
Our simulation with  ${\rm P_{\rm mn}}=5$  converged to an {average absolute vertical magnetic field strength of about 30 G at the photosphere (after 230 min simulation time) 
 and small-scale magnetic structures of surface fields of up to 930 G evolved (see Fig.\ \ref{fig:Bz_phot}), nevertheless, no magnetic bright points  were found in bolometric intensity maps, as was also reported by V{\"o}gler \& Sch{\"u}ssler (2007).
The reason for this behavior might be that in  these  mixed-polarity simulations magnetic features of high vertical field strength usually are surrounded by the opposite-polarity field, and along with the ongoing magnetic field compression process, part of the magnetic flux cancels out. This means that magnetic structures of high field strength remain too small in extent and short-lived to be seen as bright features.
In contrast, in unipolar simulations of the same type with the same average magnetic field strength, strong magnetic elements were detected after about 10min of simulation time, starting from a homogenous background field (Thaler $\&$ Spruit 2014). %
 A small-scale surface dynamo process, if it  takes place on the Sun, is  most likely 
not responsible for most of the intrinsically strong field component. It
would also be unable to explain small-scale field regions with a
statistically significant dominance of one polarity over the other. 
\subsection{Role of an imposed weak net flux}
\label{weakf}
If the experience with the shearing boxes studied in accretion physics is an indication, the presence of even a relatively weak mean flux density in the simulation may have a significant effect on a dynamo process. For the Sun, remnants of active regions spreading across the surface might have a similar effect of lowering the threshold for dynamo action. The behavior within the limit of low   ${\rm P_{\rm m}}$  and high ${\rm R_m}$ at zero mean flux would then be somewhat academic for the solar case. Simulations that address small-scale dynamo action in the presence of a dispersed strong-field component might then be more relevant.
 If it were the case that for solar conditions (${\rm R_m} \sim 10^{7}, {\rm P_{\rm m}}  \sim 10^{-5}$) 
  dynamo behavior  is also affected by the presence of weak fields,
 the strength of small-scale dynamo activity might be intermittent,
 depending on the distribution of the detritus from active regions rather
 than  being present around each and every granule on the surface, as a local convective dynamo action would predict.
\section*{Acknowledgements}{The research leading to these results has received funding from the European Research Council
under the European Union’s Seventh Framework Program (FP/2007-2013)/ERC Grant Agreement no. 307117.}


\end{document}